\def\d#1/d#2{ {\partial #1\over\partial #2} }




\newcount\eqnumber
\def\beq{ \global\advance\eqnumber by 1 $$ }
\def\eeq{ \eqno(\the\eqnumber)$$ }
\def\n{\global\advance \eqnumber by 1\eqno(\the\eqnumber)}
\def\puteqno{
\global\advance \eqnumber by 1 (\the\eqnumber)}


\def\ifundefined#1{\expandafter\ifx\csname
#1\endcsname\relax}
 \newcount\sectnumber \sectnumber=0
\def\sect#1{ \advance \sectnumber by 1 {\it \the \sectnumber. #1} }

\newcount\refno \refno=0  
\def\[#1]{
\ifundefined{#1}
\advance\refno by 1
\expandafter\edef\csname #1\endcsname{\the\refno}\fi[\csname
#1\endcsname]}
\def\refis#1{\noindent\csname #1\endcsname. }

\def\label#1{
\ifundefined{#1}
\expandafter\edef\csname #1\endcsname{\the\eqnumber}
\else\message{label #1 already in use}
\fi{}}
\def\(#1){(\csname #1\endcsname)}
\def\eqn#1{(\csname #1\endcsname)}

\baselineskip=20pt
\magnification=1200
\hoffset = 1in
\voffset = 1.25in
\hsize = 4.7in
\vsize = 7in
\def\BEGINIGNORE#1ENDIGNORE{}


\hfuzz = 0.5in
\def\label#1{
\ifundefined{#1}
\expandafter\edef\csname #1\endcsname{\the\eqnumber}
\else\message{label #1 already in use}
\fi{}}
\def\(#1){(\csname #1\endcsname)}
\def\eqn#1{(\csname #1\endcsname)}

\centerline{\bf RENORMALIZED CONTACT POTENTIAL IN TWO DIMENSIONS}
\vskip 1.5pc
\centerline{R.J. HENDERSON and S. G. RAJEEV}

 \centerline{\it Department of Physics and Astronomy, University of
Rochester, Rochester, NY 14627}

\vskip 1pc

{\narrower \baselineskip = 10pt \tenrm
We obtain for the attractive Dirac $\delta$-function potential in
two-dimensional
quantum mechanics a renormalized formulation that avoids reference to a cutoff
and running coupling constant. Dimensional transmutation is carried out
before attempting to solve the system, and leads to an interesting eigenvalue
problem in $N-2$ degrees of freedom (in the center of momentum frame) when
there are $N$ particles.
The effective Hamiltonian for $N-2$ particles has a nonlocal attractive
interaction, and
the Schrodinger equation becomes an eigenvalue problem for the logarithm
of this Hamiltonian. The 3-body case is examined in detail, and in this case a
variational estimate of the ground-state energy is given.\par}

\vskip 1.5pc
\noindent{\bf I. \hskip 0.2pc Introduction}
\vskip 1pc
\noindent
Scale invariance, and the ultraviolet divergences for which it is responsible,
is an essential feature of the quantum field theories, QCD and electroweak,
that comprise the standard model of elementary particle interactions. The
divergences that are obtained in the process of quantizing the
classical field theories on which the quantum dynamics are based can be removed
via the mathematical procedure of renormalization. In condensed matter physics
renormalization techniques are used to obtain mathematical models of physical
phenomena (e.g. phase transitions) which, despite the
presence of a characteristic scale, provided by a lattice spacing, are
scale independent in nature. In contrast, renormalization in the context of
elementary particle interactions is necessary to
make the fundamental theory physically sensible.

That an awkward renormalization procedure is necessary to permit the
quantum field theory paradigm to successfully describe elementary particle
interactions might be viewed as evidence that quantum field theory is not the
proper framework for this problem. Some other exotic and finite theory might
be more conceptually accurate and less mathematically cumbersome.
On the other hand, one may take Wilson's point of view,\[wilson],
on the role of renormalization in particle physics: a renormalizable
quantum field theory may be viewed as an {\it effective model}
which approximates for low enough energies (or long enough
distances) a more fundamental and comprehensive theory. In addition, then, to
seeking a more fundamental theory underlying renormalizable quantum field
theories, we may aspire to a deeper understanding of the very successful
effective theories we have.

Toward this end we take the point of view that renormalizable
interactions might be given a {\it finite formulation},\[hendersonc], which
avoids the need for
renormalization altogether, without the necessity of discarding the framework
of quantum
field theory. The {\it renormalized interactions} that would be part of this
formulation could not be scale-invariant; the scaling symmetry would be broken
explicitly at the outset, and not through the renormalization procedure. Other
properties of the erstwhile ``fundamental'' interactions might be
modified as well.

This approach to finding a finite, effective theory of particle
interactions would be tantamount to a reordering of the conventional analysis.
It would require renormalizing the theory completely {\it before} embarking on
efforts to solve it, rather than renormalizing in parallel with the finding
of solutions. No nonphysical cutoffs or running parameters would appear in the
formulation or solutions of the renormalized theory. Rather, such a theory
could be formulated as a well-posed mathematical
problem: a set of differential equations with appropriate boundary conditions
for example. If this point of view is the correct one, then the reason we
cannot write down such a formulation of QCD or electroweak theory is not
because they are only low-energy effective theories (though they may be) but
rather because we have not yet achieved a deep
enough understanding of these theories to write them down in the simplest way.

Although we believe finding a finite formulation of renormalizable interactions
to be a worthwhile goal,
attacking this problem directly, in say QCD, appears too formidable at
this time. We have in the past, though, found examples of simpler,
asymptotically free
renormalizable theories that lend themselves to this approach. In\[hendersona]
we renormalized the large-$N$ limit of the 1+1 dimensional non-Abelian Thirring
(or Gross-Neveu) model before finding some exact solutions. The scale-breaking
renormalized interaction manifested itself as a restriction on the domain of
the Hamiltonian operator. A similar role for the Hamiltonian domain was found
in\[hendersonb] in the study of a quantum mechanical system of
two particles in two dimensions attracted by a Dirac $\delta$-function
potential. The same system was also examined in the path integral
picture, where it was found that the renormalized interaction appeared as a
subtle modification to the Wiener measure.

In this paper we continue with our investigation of quantum mechanical
particles interacting through an attractive Dirac $\delta$-function potential
in two dimensions. We first work out the case of three particles in detail, and
eventually extend our ideas to the N-body case.

\vfill
\break

\vskip 1.5pc
\noindent{\bf II. \hskip 0.2pc Review of the Two-Body Problem}
\vskip 1pc
\noindent

Before embarking in the next section on the three-body problem, let us recall
the renormalized formulation, in the Hamiltonian picture, of the two-body
problem. In the two-body case, after separating out
the center of mass coordinate, the original Schrodinger equation in
configuration space is:

\beq
-2\Delta\Psi_{\lambda}(\bar x)
- g\delta^2(\bar x)\Psi_{\lambda}(\bar x) = \lambda \Psi_{\lambda}(\bar x)
\eeq

\noindent where $\Delta$ is the two-dimensional Laplacian, and $g$ is a
positive,
dimensionless coupling constant. We have taken the masses of the two particles
to be $m_1=m_2=1/2$ and chosen units such that $\hbar = 1$. In momentum space
the Schrodinger equation reads:
\beq
2p^2 \Psi_{\lambda}(\bar p) - {g\over {(2\pi)^2}}\int d^2p \Psi_{\lambda}
(\bar p) = \lambda \Psi_{\lambda}(\bar p)
\eeq\label{schro2}

This eigenvalue problem, however, is nonphysical since, due to scale
invariance,
the presence of even a single negative energy solution implies a continuum of
negative energy states extending down to minus infinity. On the other hand, if
one attempts to restrict the domain of the Hamiltonian to the positive energy
sector only, one finds that a complete set of eigenstates cannot be found:
there will be no zero angular momentum eigenstate, \[hendersonc]. The
Hamiltonian cannot be self-adjoint under such a choice of domain.
(The situation is analogous
to the presence of solutions to the Dirac equation of unbounded negative
energy,  and the impossibility of ignoring or "throwing away" these states).

This system's illness can be cured via
renormalization. First one regularizes the system by introducing a
momentum cutoff (upper bound). Then
the coupling constant, $g$, is required to depend on the cutoff in such a way
that the ground-state energy of the regularized system remains finite as the
cutoff is removed, i.e. taken to
infinity. This procedure removes $g$ from the problem, replacing it with a
parameter having dimensions of energy, a trade sometimes called 'dimensional
transmutation'. This new parameter is arbitrary,
characterizes the strength of the interaction, and can be taken to be the
ground-state energy of the two-body system.

The customary way to obtain renormalized solutions to the problem,
(see\[thorn],\[huang],\[gupta],\[manuel],\[amelino]), then, is to solve the
regularized system first, and then take
the limits of solutions (wavefunctions, scattering amplitudes, etc.) as the
cutoff is taken to infinity. In\[hendersonb] we showed that an equivalent, but
simpler, formulation is given by the following two equations:

\beq
2p^2 \Psi_{\lambda}(\bar p) -
\lim_{p \rightarrow \infty} 2p^2 \Psi_{\lambda}(\bar p) =
\lambda \Psi_{\lambda}(\bar p)
\eeq\label{renschro2}

\noindent and
\beq
\int d^2p (\Psi_{\lambda}(\bar p) - {{\eta_{\Psi_{\lambda}}}\over
	{2p^2 + \mu^2}})=0
\eeq\label{domain2}

\noindent where
$\eta_{\lambda}\equiv\lim_{p \rightarrow \infty}2p^2 \Psi_{\lambda}(\bar p)$.

These two equations give an example of what we mean by a {\it finite
formulation} of a renormalizable theory. The theory at this stage has been
renormalized, and can be treated as a well-posed mathematical problem. The
first equation is recognized as a renormalized version of the Schrodinger
equation. The
interaction has become the term $-\eta_{\Psi_{\lambda}} = -\lim_{p \rightarrow
\infty}2p^2 \Psi(\bar p)$ on the left hand side. This term is the
{\it renormalized interaction}. Wavefunctions for which
$\eta_{\Psi_{\lambda}}$ is zero do not take part in the interaction.
Interestingly, there
is no adjustable parameter in the Hamiltonian operator appearing in this
equation. In fact, the appropriate Hamiltonian in configuration space
is the Laplacian with no interaction term at all: the deviation from the free
theory is contained entirely in the specification of the Laplacian's domain.

The second equation specifies the domain of the Hamiltonian. In configuration
space it is a local condition that implies that wavefunctions with zero angular
momentum diverge logarithmically at the origin; in configuration space, then,
the interaction
appears as this boundary condition on wavefunctions. The parameter
$\mu^2$ has dimensions of energy, and can be picked arbitrarily. For any choice
of $\mu^2 > 0$, the Hamiltonian
will be self-adjoint (and in fact is a self-adjoint extension of the free
Hamiltonian, \[albeverio]). Choosing $\mu^2$ corresponds to selecting the
strength of
the attractive interaction and it turns out that $-\mu^2$ is the ground-state
energy, and that the free theory corresponds to letting $\mu^2 = 0$.

The above pair of equations can be solved completely, to yield all the energy
eigenstates in the two-body case,\[hendersonc] and\[hendersonb]. Although
we cannot analytically solve the N-body, or even the three-body, system, we
can formulate the problem in a similar way,
finding the analogs of equations \(renschro2) and \(domain2), which can then be
solved approximately.

\vfill
\break

\vskip 1.5pc
\noindent{\bf III. \hskip 0.2pc Renormalizing the Three-Body Problem }
\vskip 1pc

\noindent The original three-body Schrodinger
equation in configuration space is, when $m_1=m_2=m_3=1/2$ and $\hbar=1$:

\beq
(-\Delta_1-\Delta_2-\Delta_3)\Psi_{\lambda}
- g(\delta^2(\bar x_1 - \bar x_2) + \delta^2(\bar x_2 - \bar x_3)
+ \delta^2(\bar x_3 - \bar x_1)) \Psi_{\lambda} =
\lambda \Psi_{\lambda}
\eeq

\noindent where $\Delta_i$ is the two-dimensional Laplacian in coordinate
$\bar x_i$, $\Psi_{\lambda}=\Psi_{\lambda}(\bar x_1,\bar x_2,\bar x_3)$, and
$g$ is the positive, dimensionless coupling constant. In
momentum space this becomes:

\beq
(k_1^2 + k_2^2 + k_3^2) \Psi_{\lambda} -
{g\over {(2\pi)^2}}\Bigl(\int d^2k_{12} \Psi_{\lambda} + \int d^2k_{23}
\Psi_{\lambda} +
\int d^2k_{31} \Psi_{\lambda}\Bigr) =
\lambda \Psi_{\lambda}
\eeq\label{schro3}

\noindent where $\Psi_{\lambda}=\Psi_{\lambda}(\bar k_1,\bar
k_2,\bar k_3)$ and $\bar k_{ij}={1\over 2}(\bar k_i- \bar k_j)$.

As in the two-body problem, we work in momentum space, in the center of
momentum frame where $\bar K \equiv \bar k_1 + \bar k_2 + \bar k_3 = 0$.
Defining $\bar p_1 \equiv {2\over 3}({1\over 2}(\bar k_2 + \bar k_3) - \bar
k_1)$ (see \[faddeev] for a discussion of coordinate choices in the general
three-body problem), the Schrodinger equation in this frame is:

\beq
(2k_{23}^2 + {3\over 2}p_1^2) \Psi_{\lambda} -
{g\over {(2\pi)^2}}(\int d^2k_{12} \Psi_{\lambda} + \int d^2k_{23}
\Psi_{\lambda} +
\int d^2k_{31} \Psi_{\lambda}) =
\lambda \Psi_{\lambda}
\eeq\label{schro3}

\noindent where we choose the two independent coordinates to be
$(\bar k_{23},\bar p_1)$, such that $\Psi_{\lambda} =
\Psi_{\lambda}(\bar k_{23},\bar p_1)$, in which case $\bar k_{12}$ and
$\bar k_{31}$ stand for
$-{1\over 2}\bar k_{23}-{3\over 4}\bar p_1$ and $-{1\over 2}\bar k_{23}+
{3\over 4}\bar p_1$
respectively. We could just as well have chosen $(\bar k_{31},\bar p_2)$ or
$(\bar k_{12},\bar p_3)$ and would have found the same Schrodinger equation, as
$2k_{23}^2 + {3\over 2}p_1^2 = 2k_{31}^2 + {3\over 2}p_2^2 =
2k_{12}^2 + {3\over 2}p_3^2$ since the three particles are identical. From now
on we will use this symmetry to write this combination simply as $2k^2 +
{3\over 2}p^2$. The particle singled out by the choice of coordinates (e.g.
particle ``1'' with the choice $(\bar k_{23},\bar p_1)$) is sometimes called
the ``spectator'' particle. Then the other two may be referred to as the
``interacting pair'', even though in reality all three particles interact with
each other.

For the same reasons that \(schro2) was found to be nonphysical, \(schro3) also
represents an ill-posed physical problem. In this case if one tries to
restrict the Hamiltonian domain
to positive energy states, one will find that states with zero angular momentum
in the $\bar k_{23}$ variable are missing, \[hendersonc]. A renormalization of
the system is
again necessary. By carrying out the renormalization before, and independent
of,
solving the system, we can cast the problem in a form analogous to \(renschro2)
and \(domain2) above.

The first step in our renormalization program is to regularize the system by
reconsidering the problem in an artificial momentum space wherein the momenta
are bounded above by a cutoff. From Wilson's point of view,
this cutoff represents some scale beyond which we are not entitled to apply our
low-energy effective theory.
Here it is convenient to impose this
condition by requiring $2k^2 + {3\over 2}p^2 < \Lambda^2$. We must also allow
the coupling constant, $g$, to depend on $\Lambda$. By doing this we will be
able to specify that $g$ change with $\Lambda$ in such a way that the physics
predicted by the effective theory we derive is independent of the cutoff.
The resulting Schrodinger equation is:

\beq
(2k^2 + {3\over 2}p^2 - \lambda) \Psi_{\lambda}^{\Lambda} =
{g(\Lambda/\nu)\over {(2\pi)^2}} \sum_i \int^{\Lambda} d^2k_i
\Psi_{\lambda}^{\Lambda}
\eeq\label{regschro3}

\noindent where $\nu$ is an arbitrary parameter with dimensions of momentum,
$\int^{\Lambda}$ indicates the the restriction $2k^2 + {3\over 2}p^2 <
\Lambda^2$, the sum has three terms, and from now on $k_1, k_2$ and $k_3$ will
correspond to, and be used interchangeably with, $k_{23}, k_{31}$ and $k_{12}$
respectively.

The presence of $\Lambda$ breaks scale invariance, such that there is no longer
the instability problem preventing us from allowing negative energy states.
Also, if we define
$f_{\lambda,i}^{\Lambda}(\bar p_i) \equiv {g(\Lambda/\nu)\over {(2\pi)^2}}\int^
{\Lambda} d^2k_i \Psi_{\lambda}^{\Lambda}$, then \(regschro3) can be written:

\beq
(2k^2 + {3\over 2}p^2 - \lambda) \Psi_{\lambda}^{\Lambda} =
\sum_i f_{\lambda,i}^{\Lambda}(\bar p_i)
\eeq

If we decide to consider
$\Psi_{\lambda}^{\Lambda}$ to be a function of $\bar k_{23}$ and $\bar p_1$,
then it is also useful to note:

\beq
\bar p_2 \equiv {2\over 3}({1\over 2}(\bar k_3 +
\bar k_1) - \bar k_2) = -\bar k_{23} - {1\over 2}\bar p_1
\eeq\label{p2}

\noindent and

\beq
\bar p_3 \equiv {2\over 3}({1\over 2}(\bar k_1 +
\bar k_2) - \bar k_3) = \bar k_{23} - {1\over 2}\bar p_1
\eeq\label{p3}

It is left to determine what $g(\Lambda/\nu)$ must be in order that this
regularized system will have a well-defined limit as we let
$\Lambda \to \infty$. By
integrating \(regschro3) over any of the $k_i$ it is not difficult to show,
\[hendersonc], that $g(\Lambda/\nu)$ must satisfy, for large enough
$\Lambda/\nu$:

\beq
{{(2\pi)^2}\over g(\Lambda/\nu)} = \int_{k<\Lambda} {{d^2k}\over {2k^2 +
\nu^2}}
\eeq\label{g}

\noindent where we keep in mind that $\nu^2$ is arbitrary. \(g) implies that
$g(\Lambda/\nu) \to 0$ as we let $\Lambda \to \infty$, indicating that the
interaction we consider is {\it asymptotically free}.  This
expression for $g$ is, with $\nu^2$ replaced by $\mu^2$, precisely what was
found during the renormalization of the
two-body system,\[hendersonb]. This is a result we should expect. In fact it
would be have been reasonable to fix $g(\Lambda/\nu)$ using the two-body result
rather than to rederive it in the three-body case. This correspondence allows
us
to identify
$-\nu^2 = -\mu^2$ as the binding energy of the two particle bound state, i.e.
the ground-state energy of the two-body system. From now on we will use
$-\mu^2$ to denote this quantity.

By taking the $k_{23} \to \infty$ limit of \(regschro3) we can make the
identification:

\beq
\lim_{k_{23} \to \infty}2k_{23}^2\Psi_{\lambda}^{\Lambda} =
{g(\Lambda/\mu)\over {(2\pi)^2}}
\int^{\Lambda} d^2k_{23} \Psi_{\lambda}^{\Lambda}=
f_{\lambda,1}^{\Lambda}(\bar p_1)
\eeq\label{lim3}

\noindent This equation has counterparts with $\bar k_{23}$ replaced by $\bar
k_{31}$ or $\bar k_{12}$, such that \(regschro3) could also be written:

\beq
(2k^2 + {3\over 2}p^2 - \lambda) \Psi_{\lambda}^{\Lambda} =
\sum_i \lim_{k_{i} \to \infty}2k_{i}^2\Psi_{\lambda}^{\Lambda}
\eeq\label{limregschro3}

This equation does not contain $g$, and will remain valid as we let $\Lambda
\to
\infty$. In this limit, \(limregschro3) becomes the three-body renormalized
Schrodinger equation, comparable to \(renschro2) in the two-body case. Again,
no coupling parameter appears in the renormalized Hamiltonian. The parameter
characterizing the interaction strength,
which we continue to take to be the two-body binding energy, will appear in
the domain (boundary condition) equation.

We can find the domain equation by combining \(g) and \(lim3). Together, as
$\Lambda \to \infty$, these equations impose the following conditions on
wavefunctions:

\beq
\int d^2k_i (\Psi_{\lambda}(\bar k_i,\bar p_i) -
{{f_{\lambda,i}(\bar p_i)}\over
{2k_i^2 + \mu^2}})=0
\eeq\label{domain3}

\noindent where $f_{\lambda}(\bar p_i) = \lim_{k_i \to \infty} 2k_i^2
\Psi_{\lambda}$. One way of understanding the choice of $g(\Lambda/\nu) =
g(\Lambda/\mu)$ in \(g) is that $g$ must have the form which ensures that
\(lim3) will continue to hold true as we let $\Lambda \to \infty$.

\(domain3) is in fact three equations, one each for $i=1,2,3$. They are
integrals over momentum space, but in configuration space they are
local conditions dictating that the wavefunctions either go to zero or diverge
logarithmically, at a rate
determined by $\mu$, whenever the coordinates of two particles are made to
coincide. That the wavefunction can blow up under these circumstances is
unusual, but not proscribed since the singularities are square-integrable.

We thus find that the three-body analogs of the two-body equations \(domain2)
and \(renschro2) are \(domain3) and the three-body renormalized Schrodinger
equation:

\beq
(2k^2 + {3\over 2}p^2 - \lambda) \Psi_{\lambda} =
\sum_i f_{\lambda,i}(\bar p_i)
\eeq\label{renschro3}

The equations \(domain3) and \(renschro3) comprise the renormalized, finite
formulation of
the three-body problem we seek. The form of these equations indicates that the
unknown parts of the wavefunctions are essentially their limits as the relative
momenta of particles are taken to infinity, i.e. the functions $f_{\lambda,i}$,
$i=1,2,3$. We will require the particles to satisfy Bose statistics, in which
case $\Psi_{\lambda}$ must be symmetric under permutations of particle indices.
Then the $f_{\lambda,i}$ must all be the same function, which we will call
$f_{\lambda}$. $f_{\lambda}$ is a function of one momentum variable rather
than two, and so the eigenvalue problem has already been simplified.
The analogous
quantity, $\eta_{\Psi_{\lambda}}\equiv\lim_{p \rightarrow \infty}2p^2
\Psi_{\lambda}(\bar p)$, in the two-body case was simply a constant which could
be fixed by
normalization, so that the eigenvalue problem could be trivially solved. In the
N-body case we will find an eigenvalue problem in $N-2$ degrees of freedom in
the center of momentum frame.

Although the crux of our problem in solving the system is finding the functions
$f_{\lambda}$ for all eigenvalues $\lambda$, \(domain3) and \(renschro3)
contain not only $f_{\lambda}$ but the wavefunction $\Psi_{\lambda}$ as well.
It would be convenient to eliminate $\Psi_{\lambda}$ from the problem
altogether, and obtain the equation satisfied by $f_{\lambda}$ alone. This
we can do by employing the Lippmann-Schwinger formulation,\[ref], of the
Schrodinger equation. The Lippmann-Schwinger approach is conventionally
reserved
for scattering states, but here we will find it a convenient starting point in
both the scattering and bound state sectors of the theory.

The Lippmann-Schwinger equation for our three-body system can be
written down by inspection using \(renschro3). It is:

\beq
\Psi_{\lambda} =
g(\bar p_1,\bar k_{23})\delta(2k^2 + {3\over 2}p^2 - \lambda) +
{{\sum_i f_{\lambda}^{\pm}(\bar p_i)} \over
{2k^2 + {3\over 2}p^2 - \lambda {\mp}i\epsilon}}
\eeq\label{psiform3}

\noindent We choose for concreteness to write \(psiform3) in the basis
$(\bar k_{23}, \bar p_1)$.
$g(\bar p_1,\bar k_{23})$ is arbitrary, excepting the
fact that it must be symmetric under permutations of particle indices.

The energy $\epsilon > 0$ in \(psiform3) is infinitesimally small. It must
appear when
$\lambda >0$ to make the division by the singular operator $2k^2 + {3\over
2}p^2 - \lambda$ well-defined. The choice of sign, $\mp$,in the denominator
corresponds to a choice of
boundary conditions: the upper sign implying outgoing scattered waves (which is
ordinarily the physical case), and the lower sign meaning converging scattered
waves. $\epsilon$ can eventually be taken to zero, but will serve to regulate
otherwise divergent integrals in intermediate steps. We add the label $\pm$ to
$f_{\lambda}^{\pm}$ to signify this choice of boundary conditions.

The first term in \(psiform3) represents the kernel of the operator $2k^2 +
{3\over 2}p^2 - \lambda$, which appears on the left hand side of \(renschro3).
It is a solution to the {\it free} Schrodinger equation. In the case
$\lambda > 0$ it represents the unscattered portion of the wave,
and in configuration space gives the wavefunction its asymptotic behavior as
particles become infinitely separated. When $\lambda < 0$ this term disappears.
Negative energy states are comprised of only the second term, and their form is
specified by \(psiform3) up to the unknown function $f_{\lambda}$. All
information contained in \(renschro3) is also in \(psiform3).

At this point we must part ways with the usual Lippmann-Schwinger analysis. No
potential appears in \(psiform3), and we cannot use the self-consistency of
this equation alone to solve for the wavefunction. In the place of usual term
involving the potential stands the
unknown function $f_{\lambda}^{\pm}$. The only further information we can
glean about
$f_{\lambda}^{\pm}$ from \(psiform3) comes from taking the limit of this
equation as
$k_{23} \to \infty$. Since $f_{\lambda}^{\pm}(\bar p_1) = \lim_{k_{23} \to
\infty} 2k_{23}^2 \Psi_{\lambda}$, taking this limit and using \(p2)
and \(p3) gives:

\beq
\lim_{k_{23} \to \infty}(f_{\lambda}(\bar k_{23})+f_{\lambda}(-\bar k_{23}))=0
\eeq\label{flim}

However, the bosonic symmetry requires that $f_{\lambda}(-\bar k_{23}) =
f_{\lambda}(\bar k_{23})$, so that \(flim) becomes the simple asymptotic
condition:

\beq
\lim_{k_{23} \to \infty} f_{\lambda}(\bar k_{23}) = 0
\eeq\label{flima}

To gain more information about $f_{\lambda}$, we need to use \(domain3). This
once more highlights the fact that in this system, the interaction between
particles is encoded in the boundary conditions on wavefunctions rather than in
a conventional potential energy term.
Inserting \(psiform3) into \(domain3) and performing integrations where
possible yields the following implicit equation for $f_{\lambda}$:

\beq
\eqalign{
\Bigl(\ln\Bigl({1\over {\mu^2}}|\lambda - {3 \over 2}p^2|\Bigr){\mp}
i\pi \Theta(\lambda - {3\over 2}p^2)\Bigr)
&f_{\lambda}^{\pm}(\bar p)  -
{2\over \pi}\int d^2k {{f_{\lambda}^{\pm}(\bar k)} \over
{(k^2 + p^2 + \bar p \cdot \bar k - {\lambda \over 2}{\mp}
{i \epsilon\over 2})}} \cr &=
{4\over {\pi^2}} \int d^2k \enskip g(\bar
p,\bar k) \delta(2k^2-{3\over 2}p^2 - \lambda) \cr}
\eeq\label{master}

This ungainly looking equation contains all the information about
wavefunctions and energy eigenvalues we need. Imposing the requirement that
$f_{\lambda}^{\pm}$ has asymptotic behavior such that the integral in
\(master) converges will ensure \(flim) holds true. When $\lambda >0$, this
equation is an inhomogeneous linear equation in $f_{\lambda}^{\pm}$. Operating
on $f_{\lambda}^{\pm}$ on the left hand side of \(master) is a symmetric
integral operator, which must be
inverted to get $f_{\lambda}^{\pm}$ for any given $g$. When $\lambda < 0$,
the inhomogeneous term proportional to $g$ vanishes identically, leaving an
eigenvalue problem for $f_{\lambda}^{\pm}$ and $\lambda$. In the following
section we partially solve the problem of finding $f_{\lambda}^{\pm}$ for
positive and negative energies, and show
that the negative energy sector yields an explicit form for a renormalized,
nonlocal effective Hamiltonian.

\vfill
\break

\vskip 1.5pc
\noindent {\bf IV. \hskip .2pc Toward Solving the Three-Body Problem}
\vskip 1pc

\noindent All solutions to the {\it free} Schrodinger equation have positive
energy and can be categorized by the following choice of basis:

\beq
\Psi_{\lambda}^{Free} =
g(\bar p_1, \bar k_{23})\delta(2k^2 + {3\over 2}p^2 - \lambda)
\eeq\label{free}

\noindent where

\beq
g(\bar p_1, \bar k_{23}) = \sum_{\Pi}
\delta^2(\bar k_{23} - {1\over 2}(\bar \kappa_{\Pi(2)} - \bar \kappa_{\Pi(3)}))
\delta^2(\bar p_1 - (\bar \kappa_{\Pi(2)} + \bar \kappa_{\Pi(3)}))
\eeq\label{freeg}

The constant vectors $\bar \kappa_{1}, \bar \kappa_{2}$ and $\bar \kappa_{3}$
are the momenta
of the three particles and label the states. There are no restrictions on them
other than they represent the particles in the center of momentum frame wherein
$\bar \kappa_{1}+\bar \kappa_{2}+\bar \kappa_{3} = 0$. The energy is $\lambda =
\kappa_{1}^2+\kappa_{2}^2+\kappa_{3}^2$. The sum in \(freeg) is over the
six permutations
of $(1,2,3)$, and gives $g$ the bosonic symmetry we require. For our purposes
it is convenient to absorb the $k_{23}$ dependence into $p_1$ using the delta
function in \(free) and to diagonalize the angular momentum in the angle,
$\theta_{23}$, of $\bar k_{23}$. In this case $g$ becomes:

\beq
\eqalign{
g_n(\bar p_1, \theta_{23}) =& \cr e^{in\theta_{23}} \sum_{\Pi} &
e^{-in \theta_{\Pi(2),\Pi(3)}}
\delta(k_{23}^2 - {1\over 4}(\bar \kappa_{\Pi(2)} - \bar \kappa_{\Pi(3)})^2)
\delta^2(\bar p_1 - (\bar \kappa_{\Pi(2)} + \bar \kappa_{\Pi(3)}))
\cr }
\eeq\label{freegn}

\noindent where $n$ is an integer, and $\theta_{i,j}$ denotes the angle of
${1\over 2}(\bar \kappa_{i}-\bar \kappa_{j})$.

Recall that in the two-body problem only those states carrying zero angular
momentum participated in the interaction. The wavefunctions with nonzero
angular momentum were just the free ones. We find a similar situation in the
three-body case. The free wavefunctions

\beq
\Psi_{n,\lambda} =
g_n(\bar p_1, \theta_{23})\delta(2k^2 + {3\over 2}p^2 - \lambda)
\eeq\label{freesolns}

\noindent for $n \not= 0$ are the positive energy solutions we seek in the
sector of the Hilbert space where the angular momenta of any pair of particles
is nonzero. Clearly \(freesolns) solves \(renschro3) with
$f_{\lambda}=0$, for any $n$. For $n=0$, due to the integration over
$\theta_{23}$,
these wavefunctions also satisfy \(domain3) for $f_{\lambda}=0$. Note that the
right hand side of \(master) becomes zero in this case. This is consistent with
the linear operator on the left hand being nonsingular.

Free wavefunctions with quantum number $n$ not equal to zero do not scatter.
Nontrivial, i.e. interacting, wavefunctions can therefore be taken to have the
form \(psiform3),
with $g(\bar p_1, \bar k_{23}) \to g_0(\bar p_1)$. $g_0$ is the symmetrized
$(g_0(\bar p_1)=g_0(\bar p_2)=g_0(\bar p_3))$
function, for $n=0$, given in \(freegn). It is independent of $\theta_{23}$, so
the integral on the right hand side of \(master) is trivial, and this equation
becomes:

\beq
\eqalign{
\Bigl(\ln\Bigl({1\over {\mu^2}}|\lambda - {{3p^2}\over 2}|\Bigr){\mp}
i\pi \Theta(\lambda - {3\over 2}p^2)\Bigr)
&f_{\lambda}^{\pm}(\bar p)  -
{2\over \pi}\int d^2k {{f_{\lambda}^{\pm}(\bar k)} \over
{(k^2 + p^2 + \bar p \cdot \bar k - {\lambda \over 2}{\mp}
{i \epsilon\over 2})}} \cr &=
{2\over {\pi}} \Theta(\lambda - {3\over 2}p^2) g_0(\bar p) \cr}
\eeq\label{master0}

Positive energy scattering solutions, $f_{\lambda}^{\pm}$ exist for all
$\lambda >0$ and all $g_0$. Note that the operator acting on
$f_{\lambda}^{\pm}$ in \(master0) is rotationally invariant. One can let
$f_{\lambda}^{\pm}(\bar p) \sim e^{il \theta_p}$, where the integer $l$ is the
total angular momentum of the three particle system, and
solve each angular momentum sector independently. It is also worth noting that
a
rotation in the angle $\theta_{23}$ is {\it not} a symmetry of the interacting
Hamiltonian; although scattered wavefunctions have a free part independent of
this angle, the scattered components of these states will in general depend on
$\theta_{23}$.
Equation \(master0) is as far
as we will take our analysis in the positive energy sector.

For the consideration of negative energy states, let us take $\lambda =
-\eta^2$. In this case, \(psiform3) gives the form of the wavefunctions as:

\beq
\Psi_{\lambda} = {{\sum_i f_{\lambda}(\bar p_i)} \over
{2k^2 + {3\over 2}p^2 + \eta^2}}
\eeq\label{negschro3}

\noindent The infinitesimal parameter $\epsilon$ can be set to zero in this
case. When $\lambda = -\eta^2$ \(master0) becomes an eigenvalue problem
for $f_{\lambda}$ and $\lambda$:

\beq
\ln\Bigl({1\over {\mu^2}}(\eta^2 + {3 \over 2}p^2)\Bigr)
f_{\lambda}(\bar p)  -
{2\over \pi}\int d^2k {{f_{\lambda}(\bar k)} \over
{(k^2 + p^2 + \bar p \cdot \bar k + {\eta^2 \over 2})}}
= 0
\eeq\label{masterneg}

For a separable attractive potential (i.e. one having the form $<\bar k|V|\bar
k'> = -v(k)v(k')$) the Faddeev integral equation for the T-matrix, \[amato],
written as an equation for the wavefunction, can be reduced to an equation in
just one variable. Bruch and Tjon, \[bruch], have shown that, in the limit in
which such a potential is made to have zero range in real space, this equation
is precisely the eigenvalue equation we have found in \(masterneg).

Our eigenvalue problem can be brought to a more conventional form by writing
it in terms of dimensionless variables. We let $\bar p = \eta \bar x$, $\bar k
=
\eta \bar y$, and $f_{\lambda}(\bar p) = u_{\lambda}(\bar x)$. Then
\(masterneg) becomes

\beq
\ln\Bigl(1 + {3 \over 2}x^2 \Bigr)
u_{\lambda}(\bar x)  -
{2\over \pi}\int d^2y {{u_{\lambda}(\bar y)} \over
{(y^2 + x^2 + \bar x \cdot \bar y + {1 \over 2})}}
=  \ln({{\mu^2}\over {\eta^2}}) u_{\lambda}(\bar x)
\eeq\label{eigeneqn}

Denoting the operator on the left hand side of \(eigeneqn) by $W$, we
can rewrite that equation as:

\beq
W u_{\lambda} = \ln({{\mu^2}\over {\eta^2}}) u_{\lambda}
\eeq

We find, therefore, a linear eigenvalue problem for the eigenfunction
$u_{\lambda}$ and the eigenvalue $\ln({{\mu^2}\over {\eta^2}})$. Note that $W$
is a symmetric, nonlocal operator. Its eigenfunctions,
$u_{\lambda}$, give us the undetermined part (see \(negschro3)) of the
wavefunction, $\Psi_{\lambda}$.

Since the energy is $-\eta^2$, can identify the operator

\beq
H \equiv -\mu^2 e^{-W}
\eeq\label{Ham}

\noindent as the Hamiltonian, with one degree of freedom (in addition to the
center of mass) effectively integrated out, of the renormalized system in the
negative energy sector. We thus find that the logarithm of the effective
Hamiltonian of the spectator particle, in momentum space,
is an integral operator. The form of the integral operator $W$ reveals that the
renormalized, effective ``spectator Hamiltonian'' includes an attractive
potential which
is nonlocal. It is important to realize, however, that the nonlocality of this
Hamiltonian arises because we have effectively integrated out one degree of
freedom, a simplification we could make because the actual interaction is of
zero range. A positive attribute of the formulation in terms of this
one-particle Hamiltonian is that the boundary conditions, and therefore the
interactions, are built in, and do not come in through some supplementary
condition on wavefunctions.

The operator $W$ consists of two terms: an unbounded positive
kinetic-energy-like operator (a multiplication by $\ln({3\over 2}x^2 + 1)$ in
momentum space) and an interaction term (the integral operator in
\(eigeneqn)). In light of numerical and variational evidence we conjecture, but
have not proved, that the interaction part of $W$ is bounded below. Certainly
the Hamiltonian including both the ``kinetic'' and ``potential'' energy terms
is bounded below. For a proof of this see \[bruch].

If the eefective potential energy term alone is in fact bounded below it
indicates that there are a finite number of
normalizable bound states, as well as a continuum of negative energy scattering
states wherein a bound state of two particles scatters from the third particle,
and the kinetic energy of scattering is less in magnitude than the two-particle
binding energy, $-\mu^2$. These states will have counterparts in the positive
energy sector for which the kinetic energy dominates. Also in the positive
energy sector will be scattering states of three unbound fundamental particles.
In fact Bruch and Tjon, \[bruch], have carried out a numerical study of
\(eigeneqn) and concluded that there are just two three-particle bound states,
a result consistent with the above picture.

The expression for $H$ in \(Ham), in combination with the explicit form of $W$,
is one of our key results. It gives us a rare glimpse at an explicit
renormalized Hamiltonian, that in most theories can at best be asymptotically
approached order
by order in perturbation theory. In this system the entire interaction appears
in the domain of the renormalized Hamiltonian. The fact that the interaction is
of zero range allows us to obtain a compact form for the eigenvalue problem by
effectively integrating out one degree of freedom.
That the simplest way to write the resulting Hamiltonian
is in terms of its
logarithm is interesting, but should not surprise us in light of the prevalence
of logarithmic dependences in asymptotically free theories. We will find
in the N-body case as well that the Hamiltonian is best written in terms of its
logarithm. Before doing this, however, we will analyze the three-body case a
little further, finding in the next section an approximation to the ground
state and ground state energy.

\vfill
\break

\vskip 1.5pc
\noindent {\bf V. \hskip .2pc  Approximating the Three-Body Ground State}
\vskip 1pc

\noindent Our system of three particles is invariant under rotations, so that
we may use this symmetry to simultaneously diagonalize the Hamiltonian and the
total angular momentum. Letting the integer $l$ be the angular momentum quantum
number, the form of negative energy wavefunctions becomes:

\beq
\Psi_{\lambda, l} = {{\sum_i e^{il\theta_i} f_{\lambda, l}(p_i)} \over
{2k^2 + {3\over 2}p^2 + \eta^2}}
\eeq

\noindent where $\theta_i$ is the angle of $\bar p_i$.

Letting $u_{\lambda,l}(x) = f_{\lambda,l}(p)$, the eigenvalue equation,
\(eigeneqn), can be decomposed into separate angular momentum sectors, giving
us:

\beq
\eqalign{
\ln({{3}\over 2}x + 1)u_{\lambda,l}(x) -
2&\int_0^{\infty} {{dy}\over {(xy)^{l/2}}}
{{\Bigl(\sqrt{(x+y+{1\over 2})^2 - xy} - (x+y+{1\over 2})\Bigr)^l}\over
{\sqrt{(x+y+{1\over 2})^2 - xy}}}
u_{\lambda,l}(y)
\cr &= \ln({{\mu^2}\over {\eta^2}}) u_{\lambda,l}(x) \cr}
\eeq\label{evaleqn_l}

The ground state will correspond to the (or one of the) $l=0$ negative energy
state(s). Letting $-\gamma^2$ be the ground state energy, the form of this
state, which we will call $\Psi_0$, is:

\beq
\Psi_0 = {{\sum_i f_0(p_i)} \over
{2k^2 + {3\over 2}p^2 + \gamma^2}}
\eeq\label{3bodygs}

\noindent with $f_0(p)=u_0(x)$ being the eigenstate corresponding to the
lowest eigenvalue $\ln({{\mu^2}\over {\eta^2}})$ solving the equation:

\beq
\ln({{3}\over 2}x + 1) u(x) -
2\int_0^{\infty}dy {{u(y)}\over {\sqrt{(x+y+{1\over 2})^2 - xy}}}
\quad = \quad \ln({{\mu^2}\over {\eta^2}}) u(x)
\eeq\label{evaleqn_gs}

\noindent By definition $u_0$ solves this equation for the value $\eta^2 =
\gamma^2$. Equivalently we can regard $u_0$ as the function which minimizes
the quadratic functional:

\beq
Q[u] \equiv \int_0^{\infty} \ln({3\over 2}x +1) u^2(x) -
2 \int_0^{\infty}dx \int_0^{\infty}dy {{u(x)u(y)}\over
{\sqrt{(x+y+{1\over 2})^2 -xy}}}
\eeq\label{Q}

\noindent subject to the constraint $\int_0^{\infty}dx u^2(x) = 1$.

One way of obtaining an approximation to the ground state wavefunction is to
numerically solve a discretized version of \(evaleqn_gs). That is, we can
diagonalize the matrix version of the operator $W$, $W_{ij}$, obtained by
replacing the continuous variables
$0<x,y<\infty$ by discrete ones $0<i\Delta,j\Delta<N\Delta$ for integers $i,j$
and $N$ and positive $\Delta$. The task then is to diagonalize $W_{ij}$.
Doing this amounts to making a discrete approximation
to the regularized problem within which there is a high momentum cutoff.
For large
$N\Delta$ the unique positive eigenvector should give a useful picture
of the ground state of the renormalized problem.

Using this method to obtain a numerical estimate of the ground state
wavefunction shows that $u_0$ has a long, power law type tail. In fact $u_0(x)$
is
quite well approximated by a function of the form $\sqrt{b}/(b+x)$. This
rational function is normalized such that the integral of its square on the
positive real line is unity, and therefore can be taken as a variational ansatz
for $u_0$. Inserting this form in the quadratic functional $Q[u]$ and
minimizing
with respect to the parameter $b$ yields the upper bound on the ground state
energy:

\beq
\lambda_0=-\gamma^2 < -\mu^2 e^{2.6} \approx -13.5 \mu^2
\eeq

Recalling that $-\mu^2$ has the
meaning of the two-body ground state energy, we see that the three-body energy
is significantly less, which is physically reasonable. Also, this bound is
consistent with the estimate $\lambda_0 \approx -16.1 \mu^2$ found by Bruch and
Tjon using a numerical diagonlaization of the Hamiltonian. In addition they
find one other bound state, with spectator particle angular momentum $l=1$,
having energy $\approx -1.25 \mu^2$.

\vfill
\break

\vskip 1.5pc
\noindent {\bf VI. \hskip .2pc  The N-Body Problem}
\vskip 1pc

\noindent The original Schrodinger equation of the N-body problem in momentum
space is:

\beq
\sum_{i=1}^{N}k_i^2 \Psi_{\lambda} -
{g \over {(2\pi)^2}} \sum_{i<j} \int d^2k_{ij} \Psi_{\lambda}
= \lambda \Psi_{\lambda}
\eeq

\noindent where we again take $\hbar =1$ and the particle masses to be
$m_i = 1/2$. $\bar k_{ij} = {1\over 2}(\bar k_i - \bar k_j)$ are, as in the
three-body case, the relative momenta of pairs of particles. As before the
system is ill-defined due to scale invariance. We regularize the problem with a
momentum cutoff $\Lambda$ such that $\sum_{i} k_i^2 < \Lambda^2$, and define
for each pair of particles the functions:

\beq
f_{\lambda, ij}^{\Lambda} =
{{g(\Lambda/\mu)} \over {(2\pi)^2}} \int^{\Lambda} d^2k_{ij} \Psi_{\lambda}
\eeq\label{nbodyf}

\noindent $g(\Lambda/\mu)$ is as given in \(g), and the superscript $\Lambda$
on the integral indicates the restriction $\sum_i k_i^2 < \Lambda^2$. Recall
that $-\mu^2$ is the two-body ground state energy.

As $\Lambda$ is taken to $\infty$ in \(nbodyf), $g(\Lambda/\mu)$ is driven to
zero, and picks out only the logarithmically divergent part of the integral
over the relative momenta of particles. The renormalized Schrodinger equation
then becomes:

\beq
(\sum_{i=1}^{N}k_i^2 - \lambda) \Psi_{\lambda} =
\sum_{i<j} f_{\lambda, ij}
\eeq\label{renschron}

The functions $f_{\lambda,ij}$ are the limits as $\Lambda \to \infty$ of
the $f_{\lambda,ij}^{\Lambda}$, and in this limit \(nbodyf) and \(g) ensure
that $f_{\lambda,ij} = \lim_{k_{ij}\to \infty} 2k_{ij}^2 \Psi_{\lambda}$.
$f_{\lambda,ij}$ is independent, then, of $\bar k_{ij}$, and if we continue to
consider the particles to be bosons, then the necessary permutation symmetry
will imply that only one of these functions need be specified. The others will
follow by permutation of particle indices.

In a similar manner as in the two- and three-body problems, we can show that
the renormalized Schrodinger equation, \(renschron), must be supplemented by
the domain equations:

\beq
\int d^2k_{ij} (\Psi_{\lambda} - {{f_{\lambda,ij}}\over
{2k_{ij}^2 + \mu^2}})=0
\eeq\label{domainn}

There is no difficulty in using the Lippmann-Schwinger form of \(renschron)
in combination with \(domainn) to derive an equation like \(master) for the
N-body problem. The full equation is not very illuminating, however, so we do
not reproduce it here. Rather we confine ourselves to the negative energy
sector and again let $\lambda = -\eta^2$. The wavefunctions then have the form:

\beq
\Psi_{\lambda} = {{\sum_{i<j} f_{\lambda,ij}} \over
{\sum_{i=1}^{N}k_i^2 + \eta^2}}
\eeq\label{negschron}

Inserting this form into \(domainn), taking, for example, the integration to be
over the relative momentum $\bar k_{12}$, and transforming to dimensionless
variables $\bar x_i = \bar k_i/\eta$, yields the eigenvalue equation:

\beq
\ln\Bigl(1 + {1\over 2}(\bar x_1 + \bar x_2)^2 + \sum_{i=3}^N x_i^2 \Bigr)
u_{\lambda,ij}  -
{2\over \pi}\int d^2x_{12}
{  {   {{\sum_{i<j}}\atop {(i,j)\not=(1,2)}}  u_{\lambda,ij}  } \over
{(\sum_{i=1}^N x_i^2 + 1)}  }
=  \ln({{\mu^2}\over {\eta^2}}) u_{\lambda, ij}
\eeq\label{eigeneqnn}

\noindent where $u_{\lambda,ij}(\bar x_1,...,\bar x_N) = f_{\lambda,ij}(\bar
k_1,...,\bar k_N)$. One must keep in mind that $u_{\lambda,ij}$ is independent
of ${1\over 2}(\bar x_i - \bar x_j)$, so that in the center of momentum frame
this function depends on $N-2$ two-dimensional vectors. The linear operator on
the left hand side of \(eigeneqnn) is the N-body version of the operator $W$.
We may, as in the three-body system, identify the renormalized spectator
Hamiltonian in
the negative energy sector as the exponential of this operator, i.e. $H =
-\mu^2 e^{-W}$.

We conclude that the N-body problem can be given a renormalized formulation
that is in principle no more complicated than for the three-body problem.
Of course it is much more difficult to solve, or even to estimate
its ground state and ground state energy. Our most intriguing result, however,
that the simplest renormalized formulation is in terms of the logarithm of the
Hamiltonian which is an integral operator in momentum space, continues to hold
in the negative energy sector of the N-body problem for arbitrary $N$.

We have found, therefore, for all $N$, a renormalized, i.e. {\it finite},
formulation of the system of $N$ particles interacting through an attractive
Dirac $\delta$-function potential.

An undesirable feature of our formulation is the schism between the treatment
of positive and negative energy states. States for all positive values of
energy exist, and are found by inverting a linear integral operator in momentum
space. Negative energy states, in contrast, come as solutions to an eigenvalue
problem. This
is the price we pay for combining equations \(renschro3) and \(domain3) into a
formulation of the problem in which the boundary condition on wavefunctions is
built in, and not a supplementary condition.

In a
sense, however, all the essential features of the system are included in the
negative energy sector. In this sector are contained normalizable bound states,
scattering amongst these composite particles, as well as scattering of the
composite particles with the elementary excitations of the theory.
It is possible, in fact, that as we allow $N \to \infty$, keeping the
ground state energy finite, that it is only this sector which evolves into the
resulting field theory. The field theory found this way should admit a finite
formulation along the lines presented here, and is a natural direction for
future research.

\vfill
\break

\vskip 1.5pc
\noindent {\bf \hskip .2pc  Acknowledgments}
\vskip 1pc
\noindent This work was supported in part by funds provided by the U.S.
Department of Energy under grant DE-FG02-91ER40685.

\vskip 1.5pc
\noindent {\bf References}\hfill
\vskip 1pc
\def\ni{\noindent}

\ni\wilson. K.G. Wilson, Phys. Rev. D{\bf2}, 1438 (1970); Phys. Rev. B{\bf4},
3174 and 3184 (1971); Phys. Rep. {\bf12C}, 75 (1974); Rev. Mod. Phys. {\bf55},
583 (1983).

\ni\hendersonc. R.J. Henderson, Ph.D. Dissertation, University of Rochester,
Rochester, NY (1997).

\ni\hendersona. R.J. Henderson and S.G. Rajeev, Intl. J. Mod. Phys. A{\bf10},
3765 (1995).

\ni\hendersonb. R.J. Henderson and S.G. Rajeev, to appear in J. Math. Phys.
(1997).

\ni\thorn. C. Thorn, Phys. Rev. D{\bf19}, 639 (1979).

\ni\huang. K. Huang, {\it Quarks, Leptons and Gauge Fields}, World Scientific,
Singapore (1982).

\ni\gupta. K.S. Gupta and S.G. Rajeev, Phys. Rev. D{\bf48}, 5940 (1993).

\ni\manuel. C. Manuel and R. Tarrach, Phys. Lett. B{\bf328}, 113 (1994).

\ni\amelino. G. Amelino-Camelia, Phys. Lett. B{\bf326}, 282 (1994); Phys. Rev.
D{\bf51}, 2000 (1995).

\ni\amato. R. D. Amato and J. V. Noble, Phys. Rev {\bf D5}, 1992 (1972).

\ni\bruch. L. W. Bruch and J. A. Tjon, Phys. Rev. {\bf A19}, 425 (1979).

\ni\albeverio. S. Albeverio, F. Gesztesy, R. Hoegh-Krohn, and H. Holden,
{\it Solvable Models in Quantum Mechanics}, Springer-Verlag, New York, NY
(1988).

\bye